\documentstyle[12pt]{article}
\begin{document}
\def\om{\omega}
\def\omt{\tilde{\omega}}
\def\ti{\tilde}
\def\o{\Omega}
\def\bchi{\bar\chi^i}
\def\In{{\rm Int}}
\def\ba{\bar a}
\def\w{\wedge}
\def\ep{\epsilon}
\def\k{\kappa}
\def\Tr{{\rm Tr}}
\def\tr{{\rm tr}}
\def\ST{{\rm STr}}
\def\ss{\subset}
\def\ot{\otimes}
\def\bc{{\bf C}}
\def\br{{\bf R}}
\def\de{\delta}
\def\al{\alpha}
\def\la{\langle}
\def\ra{\rangle_\hbar}
\def\G{\Gamma}
\def\th{\theta}
\def\lm{\lambda}
\def\bz{\bar z}
\def\e{\hbar}
\def\ve{\varepsilon}
\def\fe{{1\over \e}(1-{\rm e}^{-\e X^3})}
\def\ex{\js\e\bar XX}
\def\KI{\big[{8\over \e^2}({\rm cosh}\jp\e X^3 -1)+
\bar XX{\rm e}^{\jp\e X^3}\big]}

\def\jp{{1\over 2}}

\def\ppp{{1\over 2\pi}}
\def\spp{{1\over 4\pi}}
\def\js{{1\over 4}}
\def\sp{{1\over 4\pi^2}}
\def\d{\partial}
\def\dz{\partial_-}
\def\dbz{\partial_+}
\def\Dz{\partial_z}
\def\Dbz{\partial_{\bar z}}

\def\be{\begin{equation}}
\def\ee{\end{equation}}
\def\bea{\begin{eqnarray}}
\def\eea{\end{eqnarray}}
\def\D{{\cal D}}
\def\G{{\cal G}}
\def\H{{\cal H}}
\def\E{{\cal E}}
\def\C{{\cal C}}
\def\A{{\cal A}}
\def\P{{\cal P}}

\def\T{{\cal T}}
\def\F{{\cal F}}
\def\n{{1\over n}}
\def\si{\sigma}
\def\Si{\Sigma}

\def\b{\beta}
\def\ga{\gamma}
\def\gc{G^{\bf C}}
\def\Gc{\G^{\bf C}}
\def\lw{D_{LWS}}
\def\LW{\D_{lws}}
\def\st{\stackrel}
\def\0{^{01}}
\def\1{^{10}}
\def\od{\sqrt{2}}
\def\aq{[\al(\phi)]_\e}
\def\bi{\bibitem}
\begin{titlepage}
\begin{flushright}
{}~
IML 99-34\\
hep-th/9911239
\end{flushright}

\vspace{2cm}
\begin{center}
{\Large \bf The formulae  of Kontsevich and  Verlinde  from the
perspective of  the Drinfeld double  }\\
[50pt]{\small
{\bf C. Klim\v{c}\'{\i}k}
\\ ~~\\Institute de math\'ematiques de Luminy,
 \\163, Avenue de Luminy, 13288 Marseille, France}

\vspace{1cm}
\begin{abstract}
\noindent A two dimensional  gauge theory is canonically associated to every
Drinfeld double. For particular doubles, the theory turns out to be e.g.
  the ordinary Yang-Mills theory,
the G/G gauged WZNW model or the Poisson $\si$-model that
underlies  the
Kontsevich quantization formula. We calculate the  arbitrary genus partition
function of the latter.
 The result is the $q$-deformation of the ordinary  Yang-Mills
partition function in the sense that the series over the weights
is replaced by the same series over the $q$-weights. For $q$ equal to a root
of unity the series acquires the affine Weyl symmetry and its truncation
to the alcove coincides with the  Verlinde formula.

\end{abstract}
\end{center}
\end{titlepage}
\newpage
\section{Introduction} The original motivation of this article was to
elucidate a relation between the Yang-Mills theory in two dimensions
and the $G/G$ gauged WZNW model. It is known \cite{Wi,BT1} that the latter
can be understood as a sort of a nonlinear deformation of the former.
  The first main result of this work shows that
 it exists a whole moduli space of two
dimensional gauge theories which contains the both theories mentioned
above as special points.

Throughout this paper, $G$ will denote a simple compact connected and simply
connected Lie group. We shall argue that the moduli space of Yang-Mills-like
theories in two dimensions based on the group $G$ coincides with  the space
 of doubles $D(G)$
of  $G$ and  we  shall refer to  the points in this moduli space
as to  Poisson-Lie Yang-Mills  theories. We define a
double $D(G)$  of an $n$-dimensional real
 Lie group $G$ to  be any $2n$-dimensional  real Lie group $D$
(containing $G$
as its  subgroup)
 such that its Lie algebra $\D$
is equipped with an symmetric  invariant  non-degenerate  ${\bf R}$-bilinear
 form
 $\la .,.\rangle$ with respect to which the Lie algebra
$\G$ of $G$ is isotropic (i.e. $\la \G,\G\rangle =0$).

 Our second  main result  is an  observation that
the Poisson $\si$-models, corresponding to the Poisson-Lie structures
on group manifolds, are special points in our moduli space.
 More precisely, they
are  the Poisson-Lie Yang-Mills theories with vanishing coupling constant.
 A particular
example of the Poisson $\si$-model is the BF theory which indeed can be
 obtained
from the ordinary Yang-Mills theory by setting to zero the coupling
constant of the latter.

The Poisson $\si$-models
 were introduced by Ikeda \cite{I} and
Schaller
and Strobl  \cite{SchS}    for manifolds without boundary. Their actions
read
\be S=\int_{\Sigma} ( C_i\w  dX^i +\jp\al^{ij}(X)C_i\w C_j).\ee
Here $X^i$ is the set of coordinates on the manifold $M$ viewed as
functions on the world-sheet $\Si$,
  $\al^{ij}(X)$ denotes a bivector field (i.e. a section of $\w^2TM$)
which defines
the Poisson structure on  $M$ and
$C_i$ is a set of $1$-form fields on $\Si$
which can
be interpreted as sections of the bundle $X^*(T^*M)\otimes T^*\Si$.

 Recently,
Cattaneo and Felder \cite{CF}  have shown that certain correlators of the
Poisson $\si$-models,
corresponding to insertions at the boundary of the $disc$, are computed
by the Kontsevich formula \cite{K}. This is in the sense of the perturbation
expansion in the field theoretic Planck constant $\hbar$.  In our
 picture,
this Planck constant $\hbar$ turns out to be  the parameter which multiplies
 the
Poisson-Lie
bracket on $G$ and therefore it can be interpreted as the Planck constant
also from this point of view.

The third main result of this paper is the computation of an arbitrary
genus partition
function of the Poisson-Lie Yang-Mills theory corresponding
to the Lu-Weinstein-Soibelman (LWS)
Drinfeld double $D(G)$. This model possesses the gauge symmetry based on
 the group $G$. For the special case of the vanishing coupling constant,
we thus obtain the partition function of  the
 Poisson $\si$-model for the LWS Poisson-Lie
 structure on the dual group $\ti G$.
If we set $q=e^{2\pi\hbar B(\psi,\psi)}$, where $B(.,.)$ is
the Killing-Cartan form on $\G$ and $\psi$ is the longest root, this
partition function $Z(q)$ has an interesting behaviour in the $complex$
plane $q$. In fact, for $q=1$ it gives the ordinary BF partition
function and for $q\neq 1$ but equal to a root of unity it gives the
standard Verlinde formula \cite{V}. It therefore appears  natural
 to refer to  the
partition function of the Poisson $\si$-model for an
 arbitrary double $D(G)$ as to
a generalized
$D(G)$ Verlinde formula.

All those results suggest that the Kontsevich and the Verlinde formulae are
in fact cousins; a general
correlator of the Poisson $\si$-model
with bulk and boundary insertions, on arbitraty genus and
for arbitrary double $D(G)$ is the object that appears to
 generalize
both of them.

In section 2, we shall define the Poisson-Lie Yang-Mills
theory as the gauge
theory canonically associated to every  double $D(G)$ and we shall indicate
the doubles  which give respectively the ordinary Yang-Mills theory and
the $G/G$ gauged WZNW model. In section 3, we review the definition of the
LWS double, we identify
 its Poisson-Lie Yang-Mills theory and show how the
Poisson $\si$-model (1) emerges
if the coupling constant vanishes. We calculate the corresponding
partition function
in section 4. We use an appropriate generalization of the method
of Blau and Thompson  worked out for the ordinary Yang-Mills theory
and for the $G/G$ gauged WZNW model in
\cite{BT1,BT2}. We shall finish with a short outlook.

\section{The Poisson-Lie Yang-Mills theory}

The Poisson-Lie Yang-Mills  theory, that  we shall associate to every
double
of a Lie group $G$,  is simply obtained by an isotropic gauging
of the WZNW model on the double. Its action reads
$$ S(l,\eta^{10},\eta^{01})=$$\be I(l)+\ppp\int_{\Si}
[\la \dbz ll^{-1},\eta^{10}\rangle -
\la \eta^{01},l^{-1}\dz l\rangle -\la l\eta^{01},l^{-1},\eta^{10}\rangle]+
\ve_i\int_{\Si}\omega O_i(l),\ee
where $l$ is a map from the world-sheet into the double $D$,
 and $\eta^{10}$
and $\eta^{01}$ are respectively $(1,0)$ and $(0,1)$
forms on the world-sheet
with values in the isotropic subalgebra $\G$ of the Lie algebra $\D$ of $D$.
$\ve_i$ are  coupling constants, $\omega$ is a volume form
on the world-sheet and  $O_i(l)$ are functions on the group
manifold $D$ which are separately invariant with respect to the left and
right action of $G$ on $D$. $\dz$ and $\dbz$ denote the Minkowski
version of the Dolbeault coboundary operators, in particular, acting on
functions they are given by the standard light-cone derivatives
\be \partial_{\pm}=\partial_{\tau}\pm\partial_{\si}.\ee
Note that the (2) is written entirely in the language of the differential
forms, though we have suppressed the symbol of the wedge product.
The pure WZNW model action is given in the standard way
\be I(l)=\spp\int \la \dbz ll^{-1},\dz ll^{-1}\rangle+
{1\over 24\pi}\int d^{-1}\la
dll^{-1},
[dll^{-1},dll^{-1}]\rangle.\ee
Of course, it is obvious how to  make sense of (4) also
when $D$ is not a matrix group.

In this section and in the following one the Poisson-Lie Yang-Mills
theory (2)
will be considered only at the classical level;
the world-sheet $\Si$ will be
always a cylinder equipped with the $Minkowski$ metric. This means, that
the variation problem with the fixed boundary
conditions at the initial and final time  is well-defined
 (see \cite{KS3} for
more details about the WZNW model on the cylinder).
When we fix a gauge, we shall always
tacitly assume that those initial and final boundary conditions are
compatible with this fixing.

The model (2) is gauge invariant with respect to two mutually commuting
gauge symmetries:
\be l\to glk^{-1}, \quad \eta^{10}\to g\eta^{10} g^{-1}-\dz gg^{-1},\quad
\eta^{01}\to k\eta^{01} k^{-1}-\dz kk^{-1},\ee
where $g,k\in G$ are mappings from the world-sheet into the maximally
isotropic subgroup
$G$. The crucial property of the action (4), which is needed for verifying
the gauge invariance (5), is the validity of the
 Polyakov-Wiegmann formula
\cite{PW}:
\be I(l_1l_2)=I(l_1)+I(l_2)+
\ppp\int\la \dbz l_1l_1^{-1},l_2^{-1}\dz l_2\rangle
\ee

Let us consider a compact group $G$ and for its double $D(G)$
we take its cotangent bundle $T^*G$. This bundle is of course trivializable,
hence we can represent every point in its total space as a pair
 $(g,X)$, where $g\in G$ and $X\in\G^*$. The group law is then
\be (g_1,X_1)(g_2,X_2)=(g_1g_2, Coad_{g_1}X_2 +X_1)\ee
and the  Lie algebra of $T^*G$ is the semidirect sum of $\G$ and $\G^*$,
where $\G$ acts on $\G^*$ in the coadjoint way. Finally, the invariant
nondegenerate bilinear form $\la .,.\rangle$ on $\D$ is given by
\be \la (\al,X),(\beta,Y)\rangle\equiv  Y(\al)+X(\beta),
\quad \al,\beta\in\G,
\quad X,Y\in\G^*.\ee

If we now partially fix the gauge by setting
$l=(1,X)$, we obtain from (2) the following
theory
\be S=-\ppp\int_{\Si} X(d\eta+\eta\w\eta)
+\ve_i\int_\Si\omega O_i(X),\ee
where $\eta=\eta\1 +\eta\0$.
 If we set all $\ve_i$, but $\ve_1$,
to zero and choose \be O_1(X)=B(X,X)\ee (we denote by the same symbol the
 dual of the Killing-Cartan form $B(.,.)$ ), we find that
(9) is nothing but the action of the standard two dimensional Yang-Mills
theory. Moreover, if $\ve_1$ vanishes, we obtain the $BF$ theory. In both
cases, the theory possesses the standard gauge symmetry with respect to the
 group $G$
and in our picture it is just the residual gauge symmetry (5) that
respects the gauge condition $l=(1,X)$. It is given by
\be X\to Coad_kX,\quad \eta^{10}\to k\eta^{10} k^{-1}-\dz kk^{-1},
\quad \eta^{01}\to
k\eta^{01} k^{-1}-\dbz kk^{-1}.\ee

An important example is given by another double of the same group $G$.
We simply
take $D(G)=G\times G$ and the invariant form $\la .,.\rangle$ on its Lie
algebra $\D\equiv\G\oplus\G$ is given by
\be \la (\al_1,\al_2),(\beta_1,\beta_2)\rangle
\equiv B(\al_1,\beta_1)-B(\al_2,
\beta_2),\ee
where $B(.,.)$ is the standard Killing-Cartan form on $\G$.
Clearly, the diagonal embedding of $G$ into $G\times G$ is isotropic, i.e.
\be \la (\al,\al),(\beta,\beta)\rangle=0.\ee
Now we fix the gauge $l=(g,1)$ and evaluate the action of our
 Poisson-Lie Yang-Mills theory
(2). We obtain
$$ S=I_B(g)+$$\be\ppp\int_{\Si}[B( \dbz gg^{-1},\eta^{10}) -
B (\eta^{01}, g^{-1}\dz g) -B(g\eta^{01}g^{-1},\eta^{10})+B(\eta^{01},
\eta^{10})],\ee
where \be I_B(g)= \spp\int  B(\dbz gg^{-1},\dz gg^{-1})+{1\over 24\pi}
\int d^{-1}B(
dgg^{-1},
[dgg^{-1},dgg^{-1}])\ee
and we have set all coupling constants $\ve_i$ to zero. Needless to say, the
model thus obtained is the gauged $G/G$ WZNW model with respect to the
 form $B(.,.)$.
The residual  gauge symmetry (5) which preserves the gauge $l=(g,1)$ is now
\be g\to kgk^{-1},\quad \eta^{10}\to k\eta^{10} k^{-1}-\dz kk^{-1},\quad
\eta^{01}\to
k\eta^{01} k^{-1}-\dbz kk^{-1}.\ee
Of course, this is the standard gauge symmetry of the $G/G$ WZNW model.

\section{Lu-Weinstein-Soibelman doubles}
The previous two choices of the doubles of  $G$ have lead
to the well-known gauge theories in two dimensions. We shall now consider
another important choice of $D(G)$, where the double of the simple,
 compact, connected
and simply connected group $G$ is the so called Lu-Weinstein-Soibelman Drinfeld
double $D_{LW}(G)$ \cite{LW}.
We shall see that this choice will  lead to the  theories known as the
Poisson-$\si$-models \cite{I,SchS,CF}.

Recall, that a general Drinfeld double \cite{AM}
is a $2n$-dimensional real  Lie group $D$ whose Lie
algebra $\D$
is equipped with an symmetric invariant   non-degenerate
${\bf R}$-bilinear form $\la .,.\rangle$.
Moreover, an existence of two $n$-dimensional
subgroups $G$ and $\ti G$  is required, such that their Lie algebras
$\G$ and $\ti \G$
are isotropic with respect to $\la .,.\rangle$
and $\D$ is the direct sum of the  vector spaces $\G$ and $\ti\G$.

 Clearly, any Drinfeld double $D$ containing $G$ as one of its isotropic
 subgroups
is the double $D(G)$ in the sense described in the introduction.
The converse need not
  be true,
 for we may
have a double of $G$ which is not the Drinfeld double. Inspite of this fact,
  we   refer to  the points in our moduli space
as to the Poisson-Lie Yang-Mills  theories. The reason is the following:
if the double is indeed the Drinfeld double then   Poisson-Lie brackets are
simultaneously induced respectively
on the group manifolds $G$ and $\ti G$. In particular, to every such
Poisson-Lie structure  we associate the corresponding deformation of
the ordinary Yang-Mills theory.

The LWS double is simply the
complexification (viewed
as the $real$ group)
$G^{\bf C}$ of $G$. So, for example,
the LWS
double of $SU(2)$ is  $SL(2,{\bf C})$. The invariant non-degenerate
form $\la .,.\ra$ on the Lie algebra $\LW$ of $\lw$ is given by
\be \la x,y\ra={1\over \e}{\rm Im}B(x,y),\ee
or, in other words, it is just the imaginary part of
the Killing-Cartan form
divided by a real parameter $\e$. Since $G$ is the real form of $\gc$,
clearly the imaginary part of $B(x,y)$ vanishes if $x,y\in\G$. Hence,
$G$ is indeed isotropically embedded in $\gc$. Note the presence of the
parameter $\e$ which indicates that
 we have actually in mind a one-parameter
family of doubles.

It turns out that $\gc$ is in fact the Drinfeld double, because
 $\gc$ is at the same time the double
of its another $n$-dimensional subgroup  $\ti G$ which coincides with
the so called $AN$ group in the Iwasawa decomposition of $\gc$:
\be \gc=GAN.\ee
For the groups $SL(n,{\bf C})$ the group $AN$ can be identified with
upper triangular matrices of determinant $1$
and with positive real numbers on the diagonal.
In general, the elements of $AN$ can be uniquely represented by means
of the exponential map as follows
\be \ti g={\rm e}^{\phi}{\rm exp}[\Si_{\al>0}v_\al E_\al]\equiv
 {\rm e}^{\phi}n.\ee
Here $\al$'s denote the roots of $\gc$, $v_\al$ are complex numbers
and $\phi$ is an Hermitian element\footnote{Recall that the
Hermitian element of  any complex simple Lie algebra $\Gc$
is an eigenvector of the
involution which defines the compact real form $\G$;
the corresponding eigenvalue
is $(-1)$ . The anti-Hermitian elements that span
 the compact real form are eigenvectors
of the same involution with the eigenvalue equal to $1$. For elements
of $sl(n,{\rm C})$ Lie algebra, the Hermitian element is
indeed a Hermitian
matrix in the standard sense.}
 of the Cartan subalgebra
of $\Gc$. Loosely said, $A$ is the "noncompact part" of the complex maximal
torus of $\gc$. The isotropy of the Lie algebra $\ti\G$ of  $\ti G=AN$
follows from (19); the fact that $\G$ and $\ti \G$ generate together
the Lie algebra $\D$ of the whole double  is evident from (18).

In general, a Poisson bracket $\al$
on a manifold $M$ is a smooth section of the bivector bundle on $M$ with
vanishing Schouten bracket
\be [\al,\al]_S=0.\ee Moreover, the Poisson-Lie bracket
on a group manifold $G$  has to be compatible with the group multiplication,
i.e.
\be \{\triangle F_1,\triangle F_2\}_{G\times G}=\triangle\{F_1,F_2\}_G.\ee
Here $\triangle F(g_1,g_2)=F(g_1g_2)$ is the
standard coproduct on the algebra of functions on the group manifold
and $\{.,.\}_{G\times G}$ is the product Poisson structure on $G\times G$:
$$\{F_1(x)G_1(y),F_2(x)G_2(y)\}_{G\times G=}$$
\be=\{F_1(x),F_2(x)\}_G G_1(y)G_2(y)
+F_1(x)F_2(x)\{G_1(y),G_2(y)\}_G,\ee
where $x$ and $y$ are coordinates on the first and second copy of
 $G$ respectively. Of course, we have by definition
\be \{F_1,F_2\}_G\equiv\al(dF_1,dF_2).\ee

Since the bivector bundle on the group manifold is trivializable
by the  left invariant vector fields, we loose no information about
the Poisson-Lie structure
$\al$ if we trade it for another object, namely
 a map $\Pi:G\to \w^2\G$ defined as follows
\be \Pi(g)\equiv \Pi_{ij}(g)\ T^i\otimes  T^j\equiv L_{g^{-1}*}\al_g,\ee
where $T^i$ is some basis of $\G$,
$\al_g$ is the value of the Poisson bivector $\al$ at the point $g$
of the group manifold and $L_{g^{-1}*}$ is the push-forward map with
respect to the left translation by the element $g^{-1}$.
The conditions (20) and (21) for the Poisson-Lie structure $\al$ translate
under (24)
to the following conditions for $\Pi(g)$:
\be  \Pi_{ij}(g)=-\Pi_{ji}(g),\ee
\be \Pi_{kl}(\nabla^k\Pi_{ij}+\jp f^{km}_{~~~i}\Pi_{mj}-
\jp f^{km}_{~~~j}\Pi_{mi})+cycl(l,i,j)=0\ee
and \be\Pi(gh)=\Pi(h)+{\rm Ad}_{h^{-1}}\Pi(g).\ee
Here $f^{km}_{~~j}$ are the structure constants  of $\G$ defined as
\be [T^k,T^m]=f^{km}_{~~~j}T^j\ee
and $\nabla^k$ is a differential operator acting on functions on $G$
 as follows
\be \nabla^k F(g)\equiv {d\over dt}F(g{\rm e}^{tT^k})\vert_{t=0}.\ee
Note that the condition (27) simply says that $\Pi(g)$ is a $1$-cocycle
 in the group
cohomology of $G$ with values in $\w^2\G$.

Now let us introduce an
$\e$-dependent family of the
Poisson-Lie brackets on the group manifolds $G$ and $\ti G=AN$,
which are called
the LWS Poisson-Lie structures. They are completely determined
by the adjoint representation of $\gc$. To describe them,
it is convenient to introduce a basis $T^i$ in $\G$
and its dual basis $\ti T_i$ in $\ti \G$. The duality means the following
relation
\be \la\ti T_i,T^j\ra=\delta_i^j,\ee
moreover we have
\be [\ti T_i,\ti T_j]=\ti f_{ij}^{~~~k}\ti T_k.\ee
A convenient choice of $T^i$'s and of  $\ti T_i$'s is given,  respectively,
by the set  $(E_\al+E_{-\al}),
i(E_\al-E_{-\al}), iH_i$ and its dual $-i\e E_\al,\e E_\al,\e H_i$. Here
$H_i$ is an (Hermitian)
orthonormal basis of the Cartan subalgebra $\T$ with respect to
the Killing-Cartan form and $E_\al,E_{-\al}$ are eigenvectors of $\T$
corresponding to roots $\al$. Of course, any other basis performs
equally well.
In fact, we could choose also a basis independent description.  It seems to
us, however,  that in these particular cirmumstances the work with some
 chosen
basis will positively influence the clarity of the exposition.
 Now for each $\e$, define
the following matrices
(cf. \cite{KS1})
\be \ti A_i^{~j}(\ti g)=\la \ti g^{-1}\ti T_i \ti g,T^j\ra,
\quad \ti B^{ij}(\ti g)=\la \ti g^{-1}T^i\ti g,T^j\ra,
\quad \ti g\in\ti G\ee
and
\be A^i_{~j}( g)=\la  g^{-1} T^i  g,\ti T_j\ra,
\quad  B_{ij}( g)=\la  g^{-1}\ti T_i\ g,\ti T_j\ra, \quad g\in G.\ee
It is then a simple matter to check, that the objects
\be \ti \Pi(\ti g)=\ti\Pi^{ij}(\ti g)\ti T_i\otimes \ti T_j =
\ti B^{ki}(\ti g)\ti A_k^{~j}(\ti g)\ti T_i\otimes \ti T_j\ee
and
\be \Pi(g)=\Pi_{ij}( g)T^i\otimes T^j =
B_{ki}( g) A^k_{~j}( g)T^i\otimes T^j\ee
define respectively the Poisson-Lie structures on the groups $\ti G$ and
$G$ \cite{FR}. This means, that  the conditions (25),(26) and (27) and their
dual analogues
are verified.  The bivectors
(34) and (35) are called the LWS Poisson-Lie structures.

The existence of the global decomposition (18) enables us to define
a natural left action of the group $G$ on its dual
$\ti G=AN$ which is called
the dressing action (\cite{AM,FR,ST,LW}). An element $g\in G$ acts on
 $\ti g\in\ti G$ as follows
\be \ ^{g}\ti g=\ti P(\ti g g^{-1}),\ee
where $\ti P$ is the map from $\gc$ onto $\ti G$ induced by the Iwasawa
decomposition (18). It is easy to verify that,
indeed, (36) defines an action
of $G$
i.e.
\be \ ^{(g_1g_2)}\ti g=\ ^{g_1}(\ ^{g_2}\ti g).\ee
There is a useful formula which clarifies the relation between the dressing
transformation (36) and the Poisson-Lie  structure $\ti\Pi$ on $\ti G$.
Indeed, the infinitesimal action of an element $\b=\b_iT^i\in\G$ on
a function $F(\ti g)$ is given by
\be \delta_\b F(\ti g)=\ti\Pi^{ij}(\ti g)\b_j\ti\nabla_iF(\ti g)\equiv
(\b^{\ti \sharp_{\ti g}})^i\ti\nabla_iF(\ti g).\ee
The fact that this is really an action, i.e.
\be [\delta_\b,\delta_\gamma]F(\ti g)=-\delta_{[\b,\gamma]}F(\ti g)\ee
follows from the (dual of the) Jacobi identity (26) and from
(the infinitesimal version of) the cocycle condition (27):
\be \ti\nabla_k\ti\Pi^{ij}(g)= f^{ij}_{~~~k}-\ti f_{kl}^{~~~i}
\ti\Pi^{lj}(\ti g)
+\ti f_{kl}^{~~~j}\ti\Pi^{li}(\ti g).\ee

Before proceeding further, let us study a limit $\e\to 0$. We fix a
basis $T^i$ in
$\G$. Then it is clear
that in the limit $\e\to 0$ the commutators of the dual generators
$\ti T_i\in \ti \G$ tend to zero
and $\ti G$ becomes an Abelian group isomorphic to $\G^*$.
In the same sense, the Lie
algebra $\G^{\bf C}$ becomes isomorphic to the semidirect sum
of $\G$ and $\ti\G$, where $\G$ acts on $\ti\G$ in the coadjoint way.
 The dressing
action (36) becomes the standard coadjoint action of $G$ on $\G^*$
in this limit, the Poisson-Lie structure $\Pi(g)$ vanishes (clearly,
 $\Pi(g)$  is proportional to $\e$). Moreover, the Poisson-Lie structure
$\ti\Pi(\ti g)$ becomes nothing but the standard linear Kirillov Poisson
bracket on $\G^*$. We conclude, that the limit $\e\to 0$ corresponds
to the previously considered
case of the cotangent bundle $T^*G$ as the double $D(G)$. Hence our
Poisson-Lie Yang-Mills theory (2) on the LWS double $\gc$ is
a $1$-parameter
deformation of the standard Yang-Mills theory (9).  Its action reads
$$ S(l,\eta^{10},\eta^{01})=$$\be I(l)+
\ppp\int_{\Si}[\la \dbz ll^{-1},\eta^{10}\ra -
\la \eta^{01},l^{-1}\dz l\ra -\la l\eta^{01}l^{-1},\eta^{10}\ra]+
{\ve\over 2\e^2}\int_{\Si}\omega \tr(l^\dagger l-1).\ee
Note that we have set all but one  $\ve_i$ in (2) to zero, and we have
 chosen canonically the biinvariant term $O_i(l)$
where $\tr$ is the trace in the
adjoint representation. With an isotropic
gauge choice
\be l=\ti P(l) =\ti g\in\ti G,\ee
the  action of the Poisson-Lie Yang-Mills theory (41) for $\gc$ becomes
$$S(\ti g,\eta\1,\eta\0)=$$
\be\ppp\int [\la \dbz \ti g\ti g^{-1}, \eta\1\ra-\la
 \eta\0,\ti g^{-1}\dz \ti g\ra
-\la \eta\0,\ti g^{-1}\eta\1\ti g\ra]+{\ve\over 2\e^2}
\int\omega\tr(\ti g^\dagger \ti g-1).\ee
Moreover, if we set $\ve=0$ and  define
\be A_i\1T^i\equiv \ti A_i^{~j}(\ti g^{-1})\eta_j\1T^i, \quad
 A_i\0 T^i\equiv \eta_i\0 T^i.\ee
then  we have  (43) as
\be S(\ti g,A\1, A\0)=\ppp\int [\la \ti g^{-1}\dbz \ti g,A\1\ra -\la  A\0,
\ti g^{-1}\dz \ti g\ra
-\ti\Pi(\ti g)( A\0,A\1)],\ee
where $\ti\Pi(\ti g)$ is the LWS Poisson-Lie structure (34).
Introduce a differential form $A$,
\be A\equiv A\1 +A\0,\ee
then we can rewrite (45) as
\be S(\ti g,A)=-\ppp\int [\la A \stackrel{\w}{,}\ti g^{-1}d\ti g\ra +
\jp \ti\Pi(\ti g)(A\stackrel{\w}{,}A)].\ee
This is precisely the Poisson $\si$-model (1)
 (written in the left-invariant frame) for the Poisson-Lie
group manifold $\ti G$. Of course,
we could introduce some coordinates $X^i$ and
 write (47) directly in the form (1). We conclude that for the vanishing
coupling constant $\ve=0$, the Poisson-Lie Yang-Mills
theory gives the Poisson
$\si$-model.

The gauge fixing (42)
is only partial, the residual group of gauge symmetry (5) consists of
the dressing gauge transformations by elements $k(\xi_+,\xi_-)\in G$:
\be \ti g\to\ ^k \ti g,\ee
\be  A\0\to k A\0 k^{-1}-\dbz kk^{-1},\ee
\be A\1\to k[A\1-((A\1)^{\ti \sharp_{\ti g}}+
\ti g^{-1}\dz\ti g)^{\sharp_k}]k^{-1}-\dz k
k^{-1}.\ee
Of course, $\sharp_k$ is defined in the dual way to (38), namely for
$W=W^i\ti T_i\in \ti\G$ we have
\be W^{\sharp_k}=(W^{\sharp_k})_iT^i\equiv \Pi_{ij}(k)W^jT^i.\ee
In the limit $\e\to 0$ the Poisson-Lie structure $\Pi(k)$
vanishes and  (50) becomes the standard gauge transformation law like (11).

The reader may convince himself, that the prescription (48)-(50) defines
indeed the action of the gauge group on the triple of the fields
$(\ti g,A\1, A\0)$. She or he may also directly check the gauge invariance
of the action (43) or (47) with respect to the gauge
transformation (48)-(50).

Infinitesimal version of the transformations (48)-(50) is given by (38)
and by
\be \delta_\b  A_i\0=-\dbz\b_i -f^{jk}_{~~i}  A_j\0\b_k;\ee
\be \delta_\b A_i\1= -\dz\b_i-f^{jk}_{~~i}  A_j\1\b_k -\ti f_{ik}^{~~l}
[(\ti g^{-1}\dz \ti g)^k+\ti\Pi(\ti g)^{km}A_m\1]\b_l.\ee
The transformation (38) is the same as in \cite{CF} for the case of the
Poisson-Lie groups, but (52) and (53) are different.
This is actually an interesting issue.
Ikeda\cite{I}, Schaller$\&$Strobl \cite{SchS} and
Cattaneo$\&$Felder \cite{CF}
have remarked a gauge symmetry of the Poisson $\si$-models that closes
only on shell if the Poisson structure is not linearly dependent on
the coordinates $X^i$. The absence of the off shell closure then requires
to use the Batalin-Vilkovisky quantization.
Our gauge  symmetry  (38), which together with (49) and (50)
closes even off-shell,
acts in the same way on the Poisson manifold as the  one in \cite{CF}.
 It is due to this fact that we find plausible to conjecture
that the Kontsevich formula can be derived
by the standard Faddeev-Popov procedure
in the special case of the Poisson-Lie structures.

As an example, consider the double $SL(2,{\bf C})$ of
$SU(2)$. We choose the basis of $su(2)$  as
\be T^j={i\over 2}\si^j,\ee
where $\si^j$ are the Pauli matrices, and the dual basis of $\ti \G$ as
\be \ti T_1=\jp\e\left(\matrix{0&1\cr0&0}\right),\quad
\ti T_2=\jp\e\left(\matrix{0&-i\cr0&0}\right),
\quad \ti T_3=\js\e\left(\matrix{1&0\cr0&-1}\right).\ee
The coordinate parametrization of the group $\ti G$ is as follows
\be \ti g=
\left(\matrix{{\rm e}^{\js\e X3}&0\cr0&{\rm e}^{-\js\e X^3}}\right)
\left(\matrix{1&\jp\e (X^1-iX^2) \cr0&1}\right).\ee
With these data, the Poisson-Lie Yang-Mills theory (43) becomes
$$ S={1\over 2\pi}\int  \Sigma_{i=1}^3C_i\w dX^i$$
$$+\ppp\int\huge\{X^1C_2\w C_3+X^2C_3\w C_1+
 [\fe+\ex ]~C_1\w C_2\huge\} $$
\be + {\ve \over 2}\int d^2\xi \KI,\ee
 where we have  set
\be X\equiv X^1+iX^2,\quad \bar X\equiv X^1-iX^2.\ee
The parameter $\ve$ is the coupling constant.  Clearly,
for $\e\to 0$ we
 recover the ordinary $SU(2)$
Yang-Mills theory, because the terms in the square brackets
in the second and
 third lines of (57)
become respectively $X^3$ and $(X^3)^2 +\bar X X$ in this limit.
If the coupling constant $\ve$ vanishes, (57) gives the  Poisson $\si$-model
which is the $\e$-deformation  of the
$BF$ theory.

The infinitesimal dressing transformation on $\ti G$ is generated by the
following vector fields
\be v^1=\ti\Pi^{i1}\ti\nabla_i=[\fe -\js\e{\rm Re}(XX)]\partial_{X^2}
-X^2\partial_{X^3}+\jp\e X^1 X^2\partial_{X^1},\ee
\be v^2=\ti\Pi^{i2}\ti\nabla_i=[-\fe -\js\e{\rm Re}(XX)]\partial_{X^1}
+X^1\partial_{X^3}-\jp\e X^1 X^2\partial_{X^2},\ee
\be v^3=\ti\Pi^{i3}\ti\nabla_i=X^2\partial_{X^1}-X^1\partial_{X^2}.\ee
One can check that those vector fields leave invariant the term
$\KI$.
We do not write the gauge transformations (49) and (50) explicitely,
because the corresponding formulas are cumbersome and not  too illuminating
anyway. Their  basic ingredients are given, however, by  components
of the Poisson-Lie bivectors. They read
\be \ti\Pi (X^i)^{12}=-\fe +\js\e\bar XX,\quad \ti\Pi(X^i)^{23}=-X^1,
\quad \ti\Pi(X^i)^{31}=-X^2;\ee
\be \Pi(u,v)_{12}=\e v\bar v,\quad \Pi(u,v)_{23}=\jp\e (uv+\bar u\bar v),
\quad \Pi(u,v)_{31}=\jp i\e(\bar u\bar v-uv).\ee
Here an element $g$ of $SU(2)$ is parametrized by two
complex coordinates $u,v$
fulfilling $\bar uu+\bar vv=1$;
\be g=\left(\matrix{u &-\bar v\cr v&\bar u}\right).\ee

Much of what we said in this section about the LWS doubles
remains true in a more general situation. Actually, if the double of a Lie
group  $G$ is a Drinfeld double $D$, it follows that it exists
the dual group $\tilde{G}$
which is also the isotropic subgroup of $D$. If,
moreover, every element $l\in D$ can be unambiguously
represented as
\begin{equation}
l=k\tilde{k},\quad k\in G,\tilde{k}\in \tilde{G}
\end{equation}
and the induced map $D\to G\times \tilde{G}$ is a diffeomorphism,
 then the Poisson-Lie
structures on the groups $G$ and $\tilde{G}$ are again given
by the expressions
(34) and (35). Of course, one uses the invariant bilinear form
that corresponds to the
double in question. The Poisson-Lie Yang-Mills theory, with the gauge
group based on $G$, corresponding to the double $D$ and with the vanishing
coupling constant,
is then again given by the Poisson $\si$-model
(47) where $\tilde{\Pi }$ is the Poisson-Lie structure
on $\tilde{G}$.

\section{The partition function}

The partition function of the ordinary
Euclidean Yang-Mills theory has been computed
by many methods \cite{Wi,BT1,BT,Wi1,Ru,HS}. Here we shall calculate
this quantity
for the LWS deformation of the Yang-Mills theory introduced in the
previous section.  We use an appropriate generalization of the method
of Blau and Thompson \cite{BT1,BT2}.
\subsection{The Wick rotation}

The definition of an Euclidean version of the Poisson-Lie Yang-Mills
 theory (2)  requires
some care. The reason is the chiral gauge symmetry (5). If we naively
replace $\dz$ by $\partial_z$ and $\dbz$ by $\partial_{\bar z}$
we cannot view the elements $\Dz gg^{-1}$ and $\Dbz kk^{-1}$
as elements of $\G$ because they are actually the elements of $\G^{\bf C}$.
This  suggests that the fields $\eta^{10}$ and $\eta^{01}$ are also
independent elements of $\G^{\bf C}$. This would change,
 however, the number
of degrees of freedom of
our theory. We may try to use the  standard prescription in gauge theory,
namely, take $\eta^{10}$ and $\eta^{01}$ in $\G^{\bf C}$ and declare
$\eta^{01}$ to be an anti-Hermitian conjugate of $\eta^{10}$.
 This would balance
 the correct
number of degrees of freedom but the independent chiral
gauge transformations (5)
of $\eta^{10}$ and $\eta^{01}$ would not respect such a  constraint.

The way out of the trouble is a partial gauge fixing. We see in the examples
(9) and (14) that we can partially fix the gauge (11) or (16) in such
a way that the residual gauge symmetry acts in the same way
on $\eta^{10}$ and
$\eta^{01}$. This makes possible to take $\eta^{10}$ as the anti-Hermitian
conjugate of $\eta^{10}$ and indeed this is the standard way how the
ordinary
Yang-Mills theory and the gauged $G/G$ model are put on the Riemann surface.

Unfortunately, the gauge fixing (42) in the LWS case still leads
to the residual gauge symmetry (49)  and (50)
which acts differently on $\eta^{10}$ and $\eta^{01}$ (or, rather, on $A\1$
and $A\0$).
We can consider, however, another gauge fixing which does make possible
to define the Euclidean version of the theory\footnote{The reason
why we did not consider immediately this new gauge fixing is simple:
we wanted to make
link to the Poisson $\si$-model (1) that
underlies the Kontsevich formula and this link was explicit in the
 gauge (42).}. It uses the Cartan decomposition \cite{Ze}
of the group $\gc$ which says that every element $l\in\gc$ can
be represented as
\be l=pg,\ee
where  $g\in G$ and $p\in P$. This decomposition is  unique.
 Here our notation is standard; if we consider the set of the Hermitian
elements of $\G^{\bf C}$ (cf. footnote 1), we have  $\P\equiv i\G$ and
\be P=\exp{\P}.\ee
The exponential mapping in (67) is one-to-one.

All this  makes possible to choose conveniently the new
gauge fixing  as
\be l=\hat P(l)=p\in P,\ee
where the map $\hat P:\gc\to P$ is induced by the Cartan decomposition (xy).
The residual gauge symmetry (5) in this gauge becomes
\be p\to kpk^{-1},\quad \eta^{10}\to k\eta^{10} k^{-1}-\dz kk^{-1},\quad
 \eta^{01}\to
k\eta^{01} k^{-1}-\dbz kk^{-1}, \quad k\in G.\ee
Now it is straightforward
to write the Euclidean version of the Poisson-Lie Yang-Mills theory (41)
in the gauge (68):
$$S_E(p,\eta\1,\eta\0,\ve,\e)
 = I_E(p)+$$\be {i\over 2\pi}\int_{\Si_g}[\la \bar\partial pp^{-1},
\eta\1\ra -
\la \eta\0, p^{-1}\partial p\ra -\la p\eta\0 p^{-1},\eta\1\ra]+
{\ve\over 2\e^2}
\int_{\Si_g} \omega\tr(p^2-1),\ee
where
\be I_E(p)={i\over 4\pi}\int_{\Si_g}
 \la \bar\partial pp^{-1},\partial pp^{-1}\ra+{i\over 24\pi}
\int_{\Si_g} d^{-1}\la dpp^{-1}, [dpp^{-1},dpp^{-1}]\ra.\ee
The gauge symmetry is given by the following transformations
\be p\to kpk^{-1},\quad \eta\1\to k\eta\1 k^{-1} -\partial kk^{-1},
\quad \eta\0 \to k\eta\0 k^{-1} -\bar\partial kk^{-1}.\ee
Remark that we use in (70) and (71)  the language of differential
 forms though
we do not indicate explicitly the wedge products between the forms. The
operators $\partial$ and $\bar\partial$ are the Dolbeault coboundary
 operators
with respect to the chosen complex structure on the Riemann surface $\Si_g$
 ($g$ indicates the  genus of the surface). The forms $\eta^{10}$ and
$\eta^{01}$
are respectively the $(1,0)$ and $(0,1)$ forms in the Dolbeault complex
and the form
$\eta\1 +\eta\0$ is in  $\G\otimes  T^*\Si_g$ and
is interpreted as a connection on the (for the simply connected $G$
necessarily) trivial $G$ bundle over $\Si_g$. In particular, it means
that
\be\eta\0=-(\eta\1)^{\dagger}.\ee
In other words, $\eta\1$ is the anti-Hermitian conjugate of $\eta\0$,
where the operation $\dagger$ is  the Hermition conjugation on $\Gc$
tensored with the complex conjugation on $T^{*{\bf C}}\Si_g$.
 Note that only the term proportional to the coupling constant $\ve$
depends on the measure on the Riemann surface, which itself is normalized
as
\be \int_{\Si_g}\omega=1.\ee

The reader should avoid a pitfall in understanding the formula (70).
It  has to do  with the fact
that the LWS double $\lw$ of
the compact simple connected and simply
connected group $G$ is isomorphic to the complexification $\gc$ of $G$.
For the purpose of defining the Euclidean version of the Poisson-Lie
Yang-Mills theory,
 we have declared the $1$-forms $\eta^{10}$ and $\eta^{01}$
to be the elements of $\G^{\bf C}$, hence seemingly to be the elements
of the Lie algebra of the double. In fact,
it is indeed  correct to say that in the Euclidean version
 $\eta^{10}$ and $\eta^{01}$
are   the elements of a complexification of the Lie algebra $\G$, but
 it is $not$ correct to interpret  $\eta^{10}$ and $\eta^{01}$ as the
elements of the Lie algebra $\D_{lws}$ of the double. The solution of this
apparent paradox is that two $different$ (though mathematically
 isomorphic)
 complexifications
 of $\G$ play role here.

In order to disentangle the two different complexifications, let us work
from the very beginning
with  the real group $\lw$ as if we did not know that
it can be identified with
 the complexification of $G$. Then consider the complexified
group $\lw^{\bf C}$
and its Lie algebra $\LW^{\bf C}$. Upon the complexification of $\LW$
to  $\LW^{\bf C}$, the subalgebra $\G\ss\LW$ gets complexified to $(\Gc)'$.
We indicate by $'$ that this complexification  is not the same as the
complexification $\Gc=\LW$. In fact, the forms $\eta\1$ and $\eta\0$ are
to be understood as
the elements of $(\Gc)'$ in full agreement with the Euclidean treatment
of the coset models (cf. \cite{GK}). Of course, the invariant bilinear
form $\la .,.\ra$ on $\LW$ gets extended onto $\LW^{\bf C}$ by bilinearity
(not sesquilinearity!).

\subsection{The measure of the path integral}

We now wish to quantize the theory (70). It actually
 resembles (the gauging of) the WZNW model
 on the symmetric space $P$ as defined
in \cite{GK}. The difference is, however, that there the Killing-Cartan
form $B(.,.)$ was used while we are using
the invariant bilinear form $\la .,.\ra$\footnote{For example,
our WZNW action $I_E(a)$ vanishes for $a\in A$ which is not the
case in \cite{GK}.}. The two models have nevertheless
some common features. They  live both on the symmetric space $P$ which
is  a  contractible manifold diffeomorphic to the Euclidean space
${\bf R}^{dimG}$. This fact means that the $d^{-1}$ of the WZNW $3$-form
(based on whatever invariant nondegenerate bilinear form)
does exist globally. As the consequence,
 we do not have to extend the map $\Si_g\to P$
to a $3$-manifold, whose boundary is $\Si_g$, if we want to determine
the contribution of the WZNW-term. Hence the level of the
WZNW model (which is equal to $1/\e$ in our case) does not get quantized
and it can be an arbitrary positive real number.

We also
 note that the gauge symmetry (72) is diagonal and hence it is
not anomalous. By the way, also the chiral symmetries
(5) of the original Poisson-Lie Yang-Mills theory (2)
are not anomalous since the
Lie algebra $\G$ is isotropic. We may interpret it by saying
that also at the
quantum level the theory (70) is the gauge fixed version of the Poisson-Lie
Yang-Mills theory (41).

The partition function of the model (70) on the genus $g$
Riemann surface is given by the following path
integral.
\be Z(\ve,\e,g)={1\over Vol(G_{\Si_g})}\int( Dp D\eta\1 D\eta\0)_g
\exp{-S_E(p,\eta\1,\eta\0,\ve,\e)},\ee
where the action $S_E$ is explicitly written in (70) and $ Vol(G_{\Si_g})$
is the volume of the gauge group. It is natural to expect that
the $G$-invariant
 measures $(D\eta\1 D\eta\0)_g$ and  $(Dp)_g$ should be based
on the bilinear form $\la .,.\ra$ that underlies the model (41) or (70).
  On the other
hand, there appears an immediate trouble in using $\la .,.\ra$
for defining the measure on the fields $\eta\1$ and $\eta\0$; indeed,
these fields are isotropic with respect to $\la .,.\ra$,
hence it is not clear
how to build up a non-zero norm on the field space.

 We can circumvent the
 trouble
by borrowing some inspiration from the Lie group theory.
The standard measure on
a simple complex  group $\gc$ is only indirectly
defined by the Killing-Cartan
form $B(.,.)$ on $\Gc$. Actually,  people define another
 $G$-invariant
bilinear form \cite{Ha} as follows
\be K(X,Y)=B(X^\dagger,Y),\quad X,Y\in\Gc,\ee
where we remind that $\dagger$ means the Hermitian conjugation in $\Gc$.
Equipped with the form $K(.,.)$, the Lie algebra $\Gc$ becomes an Euclidean
space.  By left transport
of this Euclidean form from the origin
of the group manifold $\gc$ everywhere,
the Riemannian metric and, hence, the Riemannian measure on  $\gc$ is
 canonically defined. The reason\footnote{We are indebted to P. Delorme
for this explanation.} for such a
 construction
is simple. With the choice of the positive definite bilinear form $K(.,.)$,
a standartly $K$-normalized measure is defined at the same time also for
all Lie subgroups of $\gc$.

In our case, there also exists the $G$-invariant way of turning the bilinear
form $\la .,.\ra$ into a positive definite bilinear form on $\Gc$.
For this
and also for further
purposes it is convenient to fix canonically
 a real basis of the Lie algebra $\LW$:
\be \LW=Span_{\bf R}(R^\al,J^\al,K^j,r_\al,j_\al,k_j),\ee
where
\be R^\al={1\over \od}(E_\al+E_{-\al}),
\quad J^\al={-i\over \od}(E_\al-E_{-\al}),
\quad K^j={i2H_{\al_j}\over B(H_{\al_j},H_{\al_j})};\ee
\be r_\al=
{-i\e\over \od}(E_\al+E_{-\al}),
\quad j_\al={-\e\over \od}(E_\al-E_{-\al}), \quad k_j=\e H_{\lm_j}.\ee
Here our conventions and normalizations are the same as in \cite{Co}.
This means, in particular,
\be B(E_\al,E_{-\al})=-1,\quad E_\al^\dagger=-E_{-\al},\quad [E_\al,E_{-\al}]
=-H_\al;\ee \be [H,E_\al]=\al(H)E_\al, \quad \al(H)=B(H_\al,H),\ee
where $H$ is an arbitrary element of the Cartan subalgebra
$\T$,$\al_j$'s are the simple roots and $\lm_j$'s the
fundamental weights. We recognize
in $-iK^j$'s the simple coroots to which
the fundamental weights $\lm_j$'s are dual.

Note that the (anti-Hermitian)
capital generators $R^\al,J^\al,K^j$ generate the compact
real form $\G$ of $\LW$. The small generators $r_\al,j_\al,k_j$ do not
generate a Lie subalgebra of $\LW$ but they span the vector subspace
in $\LW$ that coincides with $\P$
(cf. the discussion between Eqs.(66) ,and (67)).
 The commutation relations
of the elements of the basis (xy) give rise to real structure constants
thus they define the real Lie algebra $\LW$. In this basis,
the invariant bilinear
form $\la .,.\ra$ is given  as follows
\be \la R^\al,r_\b\ra=\delta^\al_\b,\quad \la J^\al,j_\b\ra=\delta^\al_\b,
\quad \la K^i,k_j\ra=\delta^i_j,\ee
 and all other inner products
vanish. Thus we see that small generators are
in some sense dual to the capital
ones, but this decomposition does not give rise to a Manin triple because
$\P$ is not a Lie algebra.

 Consider
an ${\bf R}$-linear flip map $\th$, $\th^2=1$,
defined by changing the capital
character
into the small one in the canonical basis (77) (e.g. $\th(R^\al)=r_\al$).
Then we define the $G$-invariant positive definite
bilinear form $K_\e(.,.)$
as follows
\be K_\e(X,Y)\equiv \la \th(X),Y\ra.\ee
Note the similarity between (76) and (83); $\th$ is
the analog of $\dagger$ in (76).
 A $2n$-dimensional Euclidean volume form $d\LW$
on the Lie algebra $\LW$, originating from $K_\e(.,.)$, is given by
$$ d\LW=(\bigwedge_{\al\in\Delta_+} (dR^\al\w dJ^\al))\w
(\bigwedge_jdK^j)\w
(\bigwedge_{\al\in\Delta_+}(dr_\al\w dj_\al))\w(\bigwedge_j dk_j)$$
\be \equiv d\T^\perp \w d\T\w d\A^\perp\w d\A.\ee
Here $dR^\al,dJ^\al, dK^j,dr_\al,dj_\al,dk_j$ is by definition
the dual basis (of the dual
space of the Lie algebra $\LW$) with respect to the $K_\e$-orthonormal
basis  $R^\al,J^\al,K^j,r_\al,j_\al,k_j$.
 We use the symbol $K_\e$, because the volume
form on $\LW$ computed from $K_\e(.,.)$ differs the volume form
coming from $K(.,.)$ by a factor $c(\G)\e^{\dim\G}$,
where $c(\G)$ is an $\e$-independent
constant.

A measure  on the
gauge fields $\eta=\eta\1 +\eta\0$  coming from $K_\e(.,.)$
 is defined by an inner product on the tangent space at each point of the
connection space $\eta$:
\be (\delta\eta_1,\delta\eta_2)=\spp\int_{\Si_g}K_\e(\delta\eta_1\st{\w}{,}
\delta\eta_2).\ee
It is independent on  the point $\eta$ in the
connection space and is gauge invariant by virtue of the
$G$-invariance of the
form $K_\e$.
It can be easily checked that a measure on $\eta$ defined
by $K(.,.)$ differs
from our measure (85) by a constant independent on $\e$.
This fact will play an
important role in what follows.

By using $K_\e(.,.)$,  we have landed (up to a normalization)
 on the same volume form
as the one standardly used in  group theory.
It is  also clear  that our measure on $P$
will differ
only by the normalization factor from the standard measure on the symmetric
space $P$ \cite{He,De}. Indeed, let us define this measure in the way
useful also for further applications.

First of all we define a volume form $d\lw$ on the $2n$-dimensional
group manifold $\lw$. We do that by the left transport of the Lie algebra
volume form $d\LW$.
Thus  $(d\lw)_l$ at a point
$l\in\lw$ is now defined
as \be (d\lw)_l=L_{l^{-1}}^*d\LW,\ee
where $L_{l^{-1}}^*$ is the pull-back  of the form $d\LW$ (defined in the
unit element of the group) by the left translation
diffeomorphism. Note that the invariant
volume form $d\lw$ is thus canonically normalized
by the bilinear form $K_\e$.

The measure $dP$ on the symmetric space $P=\lw/G$ is
the most simply defined
in the following way: consider the "projection" map
$\hat P:\lw\to P$  defined
in (68).Then the integral of an arbitrary function with compact
support
$f(p)$ on $P$ is defined by the prescription
\be \int f(p)dP\equiv {1\over Vol(G)}\int (\hat P^*f)(l)d\lw,\ee
where $\hat P^*f$ is the pull-back of the function $f$ by the "projection"
map $\hat P$ and $Vol(G)$ is the volume of the compact group $G$. Of course,
the measure on the subgroup
$G\ss\lw$ is also standartly $K_\e$-normalized hence it makes sense
to speak about the volume of $G$. In fact, we can readily write the volume
form $dG$ on $G$, it is given by
\be (dG)_g=L_{g^{-1}}^*[d\T^\perp\w d\T].\ee

We finish this subsection by noting that the measure $(Dp)_g$ of the
path integral (75) is given by the Riemann surface
point-wise product of the measures (87).

\subsection{The generalized Weyl integral formula}
Our strategy for computing (75) will be similar as in \cite{BT1,BT2}.
It means that we shall first Abelianize the theory by finding
a generalized version of the Weyl integral formula
and then we shall compute the Abelian partition function in the standard
way \cite{BT1}.  The non-Abelian origin of the Abelianized
theory will be remembered in the determinants produced by the Abelianization
procedure.

It turns out that the generalization of the Weyl
integral formula, which would
work in our setting, indeed exists.
It is given in \cite{He} (p.186) and, in
 more general setting and including the normalization, in \cite{De}.
This formula is based on another form of the Cartan decomposition which says
that $any$ element $l$ of $\gc$ can be (non uniquely) written as
\be l=gak^{-1}, \quad g,k\in G,\quad a\in A.\ee
In particular, it follows from the Cartan decomposition (66)
that the elements
of $P$ can be represented as
\be p=kak^{-1}, \quad k\in G,\quad a\in A.\ee
The  ambiguity of this representation of $p$ is clearly parametrized by
the elements of the normalizer of $A$ in $G$; we denote
this group as $N_G(A)$.
Evidently, there is a normal subgroup $Z_G(A)\ss N_G(A)$ containing the
elements of $G$
which commute with $A$. This subgroup is called
the centralizer of $A$ in $G$
and in our case it coincides with the maximal torus $T$ of $G$. From the
fact that $\exp{\T}=T$ and $\exp{i\T}=A$, we conclude that the quotient
group $N_G(A)/Z_G(A)$ is nothing but the Weyl group of $\G^{\bf C}$.
Thus the decomposition (90) is unique if we view $k$ as a class in $G/T$
and $a$ as an element of $A_+$. Here  $A_+=\exp{\A_+}$ and $\A_+$ is
the fundamental domain (=the Weyl chambre) of the action of the Weyl group
on $\A=i\T$. For this unique parametrization of $P$, we can infer the
generalized Weyl integral formula \cite{He,De} which holds for the
functions satisfying \be f(p)=f(kpk^{-1}), \quad k\in G.\ee It reads
\be \int_P f(p)dP=
{{\rm Vol}(G)\over {\rm Vol}(T)}\int_{A^+} J(a)f(a)dA,\ee
where
\be J(a)=\Pi_{\al\in\Delta}{1\over 2\e}\vert a^{\al}-a^{-\al}\vert.\ee
With a parametrization \be a=\exp{\phi^jk_j}=\exp{\e\phi^jH_{\lm_j}}\equiv
\exp{\e\phi},
\quad \phi\in\A_+,\ee we have
\be J(\phi)=\Pi_{\al\in\Delta_+}{{\rm sinh}^2(\e\al(\phi))\over\e^2}.\ee
The set of all roots of $\G^{\bf C}$ is denoted as $\Delta$, the  set
of positive roots as $\Delta_+$.
 The measure $dA$ is the standard measure on $A\ss\lw$
in the sense discussed
above; it is given by the $K_\e$-normalized volume form
\be (dA)_a =L^{*}_{a^{-1}}d\A.\ee
 The volume Vol($G$) is computed with respect to the standard measure
$dG$, defined in (88), and Vol($T$) with respect to
\be (dT)_t=L^{*}_{t^{-1}}d\T,\quad t\in T.\ee
The formula (95) gives the Jacobian $J(\phi)$; for $\e=1$ it coincides
with the Jacobian in \cite{He,De}.
Note that  the limit $\e\to 0$ makes sense and it produces the Jacobian
which arises in the Weyl integral formula for
the Lie algebra $\G$ \cite{BT1}.
In what follows, we shall use a notation often used
in the world of quantum groups; i.e.
\be [x]_\e={{\rm sinh}\e x\over\e}\ee
for an arbitrary number $x$. With this notation,
the Jacobian $J(\phi)$ becomes
the product of the $q$-numbers:
\be J(\phi)=\Pi_{\al\in\Delta_+}[\al(\phi)]^2_\e.\ee

Let us now give the proof of (95). It is clear that the integral
$\int f(p)dP$ in the l.h.s. of (92) reduces to some
 integral over $A_+$ since both the
function $f(p)$ and the measure $dP$ are invariant with respect
to the conjugation by elements of $G$ (the latter fact follows from the
simultaneous
left and right $G$-invariance of $d\lw$). It is not
difficult
to find the volume form corresponding to this integration. For this,
define first  a map $\hat A_+:P\to A_+$
that  associates to every $p\in P$ the element $a\in A_+$ under
the 	Cartan decomposition (90). Clearly, the function $f(p)$ on
$P$ satisfying
(91)  is the pull-back of some function $\ti f(a)$ on $A_+$ by
the map $\hat A_+$.
 We are looking for a function $J(a)$
such that
\be \int \ti f(a) J(a) dA_+ =
{1\over {\rm Vol}(G){\rm Vol}(G/T)}
\int (\hat P^*\hat A_+^*\ti f)(l)d\lw.\ee
Here ${\rm Vol}(G/T)$ is calculated from the measure on the homogeneous
space $G/T$ defined in a similar way  as the measure on $\lw/G$ (cf.(87)).
It then follows
\be {\rm Vol}(G/T)={{\rm Vol}(G)\over {\rm Vol}(T)}.\ee
We see from (100) that
\be J(a)(dA_+)_a=i_{\{\ ^L\T^\perp_a\}}i_{\{\ ^R\G_a\}}(d\lw)_a,\ee
where the  multivector $\{\ ^R\G_a\}$ is defined as
\be \{\ ^R \G_a\} \equiv \bigwedge_{\al\in\Delta_+}
(\ ^RR^\al_a\w\ ^RJ^\al_a)\w(\bigwedge_j\ ^RK^j_a)\ee
and the  multivector $\{\ ^L\T^\perp_a\}$ as
\be \{\ ^L \T^\perp_a\} \equiv
\bigwedge_{\al\in\Delta_+}(\ ^LR^\al_a\w\ ^LJ^\al_a).\ee
Here e.g. $\ ^RJ^\al_a$
 realizes the
right action of the generator $J^\al$ of $\G$ on the group manifold
$\lw$ at the point $a$ and  the multivector $\{\ ^L\T^\perp_a\}$
corresponds to
 the left action
of $\T^\perp$ on $\lw$ at the same point $a$.
Clearly,
 $i_V\omega$ denotes the insertion of the multivector $V$ into the form
$\omega$.
 Every such a generator, say $\ ^RJ^\al_a$, can be written as
\be \ ^R J^\al_a=L_{a*}J^\al,\quad J^\al\in\G,\ee
where $L_{a*}$ is the push-forward map corresponding to the left transport.
In a similar way, we have
\be \{\ ^L \T^\perp_a\}=R_{a*}\{ \T^\perp\}=
L_{a*}(Ad_{a^{-1}}\{\T^\perp\}).\ee
Thus we immediately arrive at
 $$ J(a)(dA_+)_a=
 i_{R_{a^*}\{\T^\perp\}}i_{L_{a^*}\{\G\}}L^*_{a^{-1}}
 (d\T^\perp \w d\T \w d\A^\perp\w d\A)=$$\be = L^*_{a^{-1}}
(i_{Ad_{a^{-1}}\{\T^\perp\}}
(d\A^\perp\w d\A)).\ee
We  calculate
\be Ad_{a^{-1}}R^\al=R^\al{\rm cosh}\e\al(\phi) +
j_\al{{\rm sinh}\e\al(\phi)\over \e};\ee
\be Ad_{a^{-1}}J^\al=J^\al{\rm cosh}\e\al(\phi) -r_\al
{{\rm sinh}\e\al(\phi)\over \e}.\ee
Inserting (108)  and (109) into (107) and
taking into account (84) and (107),
it follows
\be J(\phi)=\Pi_{\al\in\Delta_+}{{\rm sinh}^2(\e\al(\phi))\over\e^2}=
\Pi_{\al\in\Delta_+}[\al(\phi)]^2_\e.\ee
\subsection{The maximal torus bundles}
Consider the gauge
\be p=a=\exp{\e\phi}, \quad \phi=\phi^jH_{\lm_j}\quad \phi\in\A_+\ee
and introduce  a pair of real ghosts  $c^r_\al,c^j_\al$
for each $positive$ root $\al$.
 By taking into account
(70), (92), (110)  and (111), the formula (75) for
the partition functions becomes
$$ Z(\ve,\e,g) ={1\over  Vol(T_{\Si_g})}
\int( D\phi D\eta\1 D\eta\0 Dc^r Dc^j)_g \exp{-\ve\int
\omega \sum_{\al\in\Delta}[\al(\phi)]^2_\e}$$
\be\exp{-{i\over 2\pi}\int \{d\phi^j\w A_j+
\sum_{\al\in\Delta_+}(\aq B_\al\w C_\al+\omega\aq^2 c^r_\al c^j_\al)\}}.\ee
Here
\be \eta\1={A\1}_mK^m+{B\1}_{\al} R^\al+{C\1}_{\al}J^\al;\ee
\be \eta\0={A\0}_mK^m+{B\0}_{\al} R^\al+{C\0}_{\al}J^\al;\ee
and the component forms $A\1_m,B\1_{\al},C\1_{\al}$
and $A\0_{m},B\0_{\al},C\1_{\al}$ are pairwise complex conjugated forms in
$T^{*{\bf C}}\Si_g$.

Note that the term $I_E(p=a)$ in (70) gives
no contribution to (112). This seems to be a
trivial fact because $\A$ is commutative and isotropic
with respect to $\la.,.\ra$.
 However, the isotropy
and commutativity  $together$ with the cohomological
triviality of the WZNW term explains the vanishing
of $I_E(a)$. In the case of the $G/G$
gauged WZNW model the WZNW term is not cohomologically
trivial and it may and does contribute \cite{BT1,Gaw1}.
This can be easily understood
by realizing that in the cohomologically nontrivial
situation one has to consider the mappings
extended to three-dimensional domain of
which $\Si_g$ is the boundary. This extension need not
respect the (isotropic and/or  commutative) gauge choice on $\Si_g$.

Our  formula (112) almost coincides with the result of the
Abelianization of the ordinary
 Yang-Mills theory (formula (2.58) of \cite{BT1}).
Up to a trivial
overall $2\pi$ normalization\footnote{Blau and Thompson
have chosen the overall
normalization to be in accord with the fixed point theorems
while we are using
the standard WZNW normalization.}, the only difference
in the action consists
in replacing the "ordinary" numbers $\al(\phi)$ of Blau and
Thompson by our quantum
numbers $[\al(\phi)]_\e$.  Moreover, the  measure of our path
integral differs from that of the
ordinary Yang-Mills case only by an $\e$-independent constant.
This constant originates from the
difference between our measure-defining form $K_\e(.,.)$ and
the ordinary Yang-Mills
 measure-defining form $K(.,.)$, where the $\e$-dependent part
of this difference
is already taken into account in the Jacobian $J(\phi)$ or,
in other words, in the ghost part of
the action.
 Moreover, also the $2\pi$ renormalization of the measure of
the gauge fields
play role (cf. our Eq.(85) and Eq.(2.11) of \cite{BT1}).
We shall eventually use a freedom  to
renormalize our measure-defining bilinear form  by all these
$\e$-independent constants in such
a way that
 the limit $\e\to 0$ in (112) gives a correctly normalized ordinary
Yang-Mills partition function.

There is a very important aspect of exploiting the
(generalized) Weyl integral formula in the
ordinary Yang-Mills case \cite{BT1} and also in our
Poisson-Lie Yang-Mills setting. It has to do with
the following fact: the validity of the Cartan
decomposition (90) of an arbitrary element $p\in P$
does not imply that the smooth mapping
$p(\bz,z):\Si_G\to P$ can be  smoothly decomposed as
\be p(\bz,z)=g(\bz,z)a(\bz,z)g(\bz,z)^{-1},
\quad g(\bz,z)\in G,\quad a(\bz,z)\in A_+.\ee
Of course,   the mappings $g(\bz,z)$ and  $a(\bz,z)$ do exist but
it is by no means guaranteed that  they   be smooth.
This fact has serious implications for the proper meaning
of the formula (112). Strictly speaking,
we $cannot$ choose the gauge (111) smoothly.
 Then what do we mean by Eq. (112)?

A  hard work of Blau and Thompson \cite{BT2}
 was needed for solving this problem in the case
of the ordinary Yang-Mills theory (and also in the case
of the $G/G$ gauged WZNW model). Fortunately, we can fully
rely on their results also in the LWS case
because $P$ is diffeomorphic to the Lie algebra
$\G$ of $G$ and this (exponential) diffeomorphism
commute with conjugations by the elements of $G$.
In fact, we have
\be P=\exp{i\G}.\ee
This means that we map  $p(\bz,z)\in P$  into $\G$ by
taking the inverse of the mapping (116) and
then we apply the Blau-Thompson diagonalization i.e.
\be -i{\rm ln}p(\bz,z)=g(\bz,z)t(\bz,z)g^{-1}(\bz,z),
\quad t(\bz,z)\in \C_+\ss\T.\ee
Here $\C_+=-i\A_+$ is the Weyl chamber in $\T$.
By multiplying (117) by $i$ and then exponentiating,
we arrive at the seeken Cartan decomposition
\be  p(\bz,z)=g(\bz,z)a(\bz,z)g^{-1}(\bz,z),
\quad a(\bz,z)=e^{it(\bz,z)}\in A_+.\ee

The analysis  \cite{BT2} of
the diagonalization of the type (117) can be translated
into our context along
the lines above and it gives the following results:
If ln$p(\bar z,z)$ is a smooth map  from the Riemann surface
$\Si_g$ into a subset of regular
elements of $\G$ then
\vskip1pc
\noindent 1) The smooth decomposition (115) can always
be  achieved locally on $\Si_g$.

\noindent 2) The diagonalized map $a(\bz,z)$ can
always be chosen to be smooth globally.

\noindent 3) Non-trivial $T$-bundles on $\Si_g$ are
the obstructions to finding smooth functions
$g(\bz,z)$ globally. In particular, if there are no
nontrivial principal  $G$-bundles on $\Si_g$
(like in our case), all isomorphism classes of torus
bundles appear as obstructions.

\noindent 4) The gauge field path integral should
include a sum over the $\T$-connections
on all isomorphism classes of $T$-bundles on $\Si_g$.
\vskip1pc
\noindent Actually, the point 4) shows in which sense
we should understand the formula (112).

 It is not so difficult to understand intuitively,
what is going on here. If the function
$g(\bz,z)$ is not smooth somewhere, then passing
from $p(\bz,z)$ to $a(\bz,z)$ is a singular
gauge transformation and the $g(\bz,z)$-transformed connection field
 $\eta$ becomes singular. It is well-known
that singular connections can be sometimes
interpreted as  connections on  nontrivial bundles
(see \cite{BT1,Kl} for examples).

The condition that $-i$ln$p(\bar z,z)$ is a regular element
of the Lie algebra ${\cal G}$ may seem inconspicuous but it
is in fact crucial for the  proper definition of the path integral.
By restricting  our space of fields $p(\bar z,z)$ to those
verifying the condition of the regularity, we make a certain choice.
We can certainly understand it simply as a part of
a plausible $definition$
of the path integral, because,  as it was shown in
\cite{BT2} such regular maps with
values in the Lie algebra $\G$ are generic. In order
to corroborate this choice
we give two arguments (a more detailed discussion
is provided in \cite{BT1,BT2} and it
is directly relevant also to our Poisson-Lie Yang-Mills case):

First of all,  the restriction to the regular maps
gives the correct answer
for the ordinary Yang-Mills case, which is confirmed by alternative
methods of calculation \cite{Wi,BT,Wi1,Ru,HS}.
Secondly, the non-regular $\phi$'s are anyway
automatically suppressed from the path integral since the
Jacobian $J(\phi)$,
 originating
from the generalized
Weyl integral formula, vanishes for them. Indeed,
 the Jacobian $J(\phi)$ vanishes if and only if $-i\phi$ is not
a regular element of ${\cal T}$. This can be seen directly from
the definition of the regularity; an element $X$
of the Cartan subalgebra
${\cal T}$ is regular iff it satisfies a condition
${\rm det}_{{\cal T}^\perp}$(ad$(X))\neq 0$ where the notation
means that the ad($X$) operator is restricted to ${\cal T^\perp}$.
Of course, if this ad($X$) determinant vanishes then there exists a
root
$\alpha$ such that $\alpha(X)=0$. The latter fact implies the
vanishing of the Jacobian $J(X)$. The opposite direction can be also
easily proved.

\vskip1pc

Following closely \cite{BT1}, we shall now perform
the path integral over the affine space of
Abelian connections $A_j$.
The maximal torus bundles are  parametrized by the
monopole numbers $(n_1,\dots,n_{rank\G})$
given by
\be \int_{\Si_g}F_j(A)=2\pi n_j,\ee
where  $A=A_jK^j$ is a connection  on the bundle
and
 $F(A)=F_j(A)K^j$ is its curvature. For each set of
the monopole numbers we choose one connection
$A^{cl}$ which we call the classical monopole solution. It fulfils
\be dA_{cl}=2\pi n_jK^j\omega .\ee
Now every connection $A$ can be written as
\be A=A_{cl}+A_q,\ee
where the "quantum part" $A_q$ of the connection is a
$1$-form on $\Si_g$ with values
in $\T$. The path integral over $A$ then
becomes the sum over the monopole numbers and the path
integral over $A_q$. The latter
imposes a constraint
\be d\phi=0.\ee
The details of this procedure which, of course, must be
accompanied by an appropriate gauge
fixing and ghost integration are presented in \cite{BT1}
Eqs. (2.71)-(2.80). Their calculation applies
to our case without any change.
Thus the integral over
$\phi$ reduces to a finite integral over the constant
mode of $\phi$ and we shall note the corresponding
measure $d\phi$ instead of $D\phi$. Of course, the integral over $d\phi$
is taken over the (interior of the) Weyl chamber $\A_+=-i\C_+$.
 The result is
$$ Z(\ve,\e,g)=\int(d\phi DB DC Dc^r Dc^j)
\exp{-\ve\sum_{\al\in\Delta}
[\al(\phi)]^2_\e}$$
\be \sum_{n_1,\dots,n_{rank\G}}\exp{\{i n_j\phi^j
-{i\over 2\pi}\int_{\Si_g}\sum_{\al\in\Delta_+}(\aq B_\al\w C_\al+
\omega\aq^2 c^r_\al c^j_\al)\}}.\ee
Here we have also used Eq. (74).

The path integral over the $B$,$C$ fields and over the ghosts
was performed in \cite{BT1} in generality,
which covers not only the ordinary Yang-Mills and the $G/G$
gauged WZNW case but also our Poisson-Lie
Yang-Mills case. The functions $M_\al$ defined in
Eq. (B.5) of \cite{BT1} are in our context
$M_\al=\aq$. We infer from (B.23) of \cite{BT1} that
\be Z(\ve,\e,g)=\int d\phi(\Pi_{\al\in\Delta_+}\aq)^{2-2g}
\sum_{n_1,\dots,n_{rank\G}}
\exp{\{i n_j\phi^j-\ve\sum_{\al\in\Delta}[\al(\phi)]^2_\e\}}.\ee
Contrary to the $G/G$ case and in accord with
the ordinary Yang-Mills theory, there is
no shift of a "level" $1/\e$.
The  last step of calculation consists in performing
the $d\phi$ integral. We use the
well-known formula
\be \ppp\sum_{n_j} e^{i n_j\phi^j}=\sum_{m^j}\delta(\phi^j-2\pi m^j).\ee
Here on the right-hand-side we recognize the periodic $\delta$ function.
The usual $2\pi$
factor in this formula is understood to be hidden
in the definition of the measure $d\phi$. Substituting
the expression (125) into (124), we arrive at
 $$ Z(\ve,\e,g)=$$\be{1\over \vert W\vert}
\sum_{m^1,\dots,m^{rank\G}}(\Pi_{\al\in\Delta_+}[2\pi m^j
\al(H_{\lm_j})]_\e)^{2-2g}
\exp{-\ve\sum_{\al\in\Delta}[2\pi m^j
\al(H_{\lm_j})]^2_\e}.\ee
Note that here we have conveniently extended the domain of
definition of $\phi$
from $\A_+$ to whole $\A$ and we compensated this by factoring
the volume $\vert W\vert$ of
the Weyl group.
 We
now interpret the summation over $m^j$ as summation
over the weight lattice of $\G$. The latter
is defined as
\be \Lambda={\bf Z}[\lm_1,\dots,\lm_r],\ee
where $r=rank\G$ and $\lm_i$ are the fundamental weights. We set
$\lm=m^j\lm_j$ and we rewrite (126) as
\be  Z(\varepsilon,\e,g)={1\over \vert W\vert}
\sum_{\lm+\rho\in\Lambda_r}\Pi_{\al\in\Delta_+}
([B(\al,\lm+\rho)]_{2\pi\e})^{2-2g}
\exp{-\varepsilon\sum_{\al\in\Delta}[B(\al,\lm+\rho)]^2_{2\pi\e}}.\ee
Note the shift by the Weyl vector $\rho=\jp\sum_{\Delta_+}\al=
\sum_j\lm_j$. Since we anyway sum up
over the whole weight lattice, this shift can be interpreted
as a pure change of the summation variable.
Another important remark concerns the notation $\Lambda_r$ in (128).
By this we mean that we sum only
over the regular points of the weight lattice
in accord with the discussion above. There is
a simple criterion to decide whether an element
of the weight lattice is regular or not. In fact,
the non-regular elements are precisely those which are
located on the walls of  the Weyl chambers.

The formula (128) is our final result for the partition function
of the LWS Poisson-Lie Yang-Mills theory, or for $\ve =0$, of the
Poisson $\si$-model corresponding to the LWS
Poisson-Lie structure on $\ti G=AN$.
 We see that, indeed, we can interpret this partition function
as the  series over the $q$-numbers. In the limit $\e\to 0$,
our result
agrees with the ordinary Yang-Mills
partition function Eq. (2.86) of \cite{BT1}.
To see this, we just have to use
 the well-known identity
\cite{Co}
\be B(x,y)=\sum_{\al\in\Delta}B(x,\al)B(\al,y),\quad x,y\in \G^*\ee
for $x=y=\lm+\rho$.

\subsection{The Verlinde formula}
Let us rewrite  our formula (128) for the partition function
$Z(\ve,\hbar,g)$ for the case
$\ve=0$:
\be  Z(\varepsilon=0,q,g)=
{1\over \vert W\vert}A^{1-g}\sum_{\lm+\rho\in\Lambda_r}\Pi_{\al\in\Delta}
\vert 1-q^{(\al,\lm+\rho)}\vert^{1-g}.\ee
Here $q=\exp{2\pi\e B(\psi,\psi)}$, $A$ is a normalization constant
to be discussed later and the bilinear form (.,.) on the dual of $\T$
is defined by the following rescaling of the Killing-Cartan form:
\be (X,Y)={2\over B(\psi,\psi)}B(X,Y).\ee
Here $\psi$ denotes the longest root.

 In our construction, $q$ was a real parameter, nevertheless,
 we shall now consider
 $Z(0,q,g)$ as a function of a $complex$ $q$. Strictly speaking,
 if $q$ is complex, we
cannot say apriori whether $Z(0,q,g)$ is a partition function
of some theory. We shall see,
however,
that for $q$ being a root of unity, $Z(0,q,g)$ can be  (almost)
interpreted as the partition
function of the $G/G$ gauged WZNW model. As it is well-known,
the latter is given by the Verlinde
formula \cite{V} which is a $finite$ $sum$ and not a series.
In which sense can we say that
our series (130) gives the Verlinde formula for $q$ being a root of unity?

The point is that for $q$ equal to a root of unity, say $q^k=1$,
the  expression under
the summation symbol in (130) acquires the affine Weyl symmetry.
The affine Weyl group \cite{PSG}
is a semidirect product of the standard Weyl group  and of
the coroot lattice.  The action of the Weyl
group on the weight lattice is standard and since
the product $\Pi_\al$ in (130) is taken over
all roots, the expression (130) is  Weyl invariant
(this follows from an idempotency and
Killing-Cartan orthogonality
of  Weyl reflections).
An element $\beta^\vee$ of the  coroot lattice acts on the weight lattice as
\be \lm_{\beta^\vee} = \lm +k\jp B(\psi,\psi) B(\beta^\vee,.)=
\lm +k(\beta^\vee ,.)^*.\ee
Here use the  symbol $(.,.)^*$ for the  form on $\T$ dual to $(.,.)$.

Recall that for a root $\beta$, the coroot $\beta^\vee\in\T$ is defined by
\be \beta^\vee={2H_\beta\over B(\beta,\beta)}.\ee
The affine Weyl symmetry of (130) for $q^k=1$ is now obvious because
\be q^{(\al,\lm_{\beta^\vee})}=q^{(\al,\lm) +k\al(\beta^\vee)}=
q^{(\al,\lm)}.\ee
The last equality follows from the fact that
$\al(\beta^\vee)$ is integer, as the result of the contraction of
$\al$ (which is also an element
of the weight lattice) with the coroot $\beta^\vee$.
Actually, for $q^k=1$, the expression (130) makes sense only if the weights
lying on  affine Weyl orbits of the non-regular weights are
excluded from $\Lambda_r$. It is precisely in this sense that we understand
(130).

We conclude that, for $q^k=1$, the series (130)
can be written as a summation over the
fundamental domain of the affine Weyl group multiplied by its (infinite)
volume. It is because of this infinite volume
renormalization that we have said above that $Z(0,q,g)$ can be $almost$
interpreted as the partition function of the $G/G$ gauged WZNW model.
  The fundamental domain of the action of the affine
Weyl group is often referred to as the
(Weyl) alcove and it contains those elements
$\lm$ of the standard Weyl chamber that fulfil
\be  \lm(\psi^\vee)< k.\ee
Here $\psi^\vee$ is the coroot of the longest root.
For $G=SU(r+1)$, the condition (135) translates
into
\be \sum_{j=1}^r m^j< k,\quad m_j>0,\ee
where $\lm=\sum_j m^j\lm_j$ are the dominant
weights sweeping the Weyl chamber.

Eq. (130)  with the summation  restricted  to the alcove is nothing
but the Verlinde
formula.
Its correct  normalization  can be achieved by
adding to the action (70) a suitable "counterterm"
of the form
$const\int_{\Si_g} R$, where $R$ is the  curvature on the Riemann surface.

As we have already remarked, it is not clear from our derivation
whether $Z(0,q,g)$ is  a partition function
of some theory if $q$ is not real. On the other hand,
the result for $q$ equal to a root of unity
suggests that it is indeed so because the partition
function of the $G/G$ gauged WZNW model
is given by the Verlinde formula. It might be that this
fact is related to an observation made
by Alekseev, Schaller and Strobl \cite{ASS}. They remarked that,
 modulo some $\delta$-function relict of
the WZNW term, the $G/G$ gauged WZNW model can be
represented as the Poisson $\si$-model where
the Poisson structure on $G$ is a complex bivector.
If $q$ and, hence, $\e$ has a non-vanishing
imaginary part then in our context the real Poisson
bivector $\ti \Pi(\ti g)$ gets
rescaled by a complex Planck constant $\e$ and it
becomes complex too. However, there is a crucial
difference between our theory and that of  Alekseev et al.
Our LWS Poisson $\si$-model
(or  the  Poisson-Lie Yang-Mills theory with the vanishing
 coupling constant)
lives on the dual group $\ti G=AN$ while the Poisson
$\si$-model of \cite{ASS} lives on the compact
group $G$. Inspite of this, the partition functions
for particular $q$'s happen to
coincide! This suggests that
 a sort of a quantum Poisson-Lie T-duality \cite{KS1,AKT}
between the two topological
$\si$-models takes place here.

\section{Outlook}

An open field for future investigations is the study of
the correlation functions of
the Poisson-Lie Yang-Mills theories. It seems to be very
plausible that, like in the ordinary Yang-Mills
case,  one can  identify observables whose
correlators are insertion independent.
It would be also interesting to calculate exactly the
three-point boundary correlator of the LWS
Poisson $\si$-model
 on the disc. It is known \cite{CF} that the perturbation
expansion in $\hbar$ gives the
Kontsevich formula. If this correlator can be computed  by a
closed formula, we might attempt
to consider a convergence  in $\hbar$. Thus $\hbar$ would make sense
not only  as the formal expansion parameter.

A more algebraic problem  consists in attempting a
classification of all doubles
 $D(G)$ of a connected simple compact group $G$. This would
lead to a generalization of the
 classification \cite{LS} of
the Poisson-Lie structures on $G$.

The Verlinde formula was recently associated to the
$quantum$ double of a finite group \cite{KSSB}.
Although we have dealt with with the $classical$
double of a Lie group, it would be certainly
worth looking at possible relations of the two constructions.

 It would be desirable to produce
a derivation of the partition function by an alternative method.
 For example, the gluing and pasting
procedure of \cite{BT}. It may seem problematic to use this
 approach  because
of problems with the definition of the WZNW action on
surfaces with boundaries. On the
other hand  we note that
in the LWS Poisson-Lie Yang-Mills
case the WZNW term is cohomologically trivial therefore this problem
 should be accessible.
Of course, this last remark is pertinent also for
the very definition of the
Poisson-Lie Yang-Mills theory on the
Riemann surfaces with boundary and for the implications concerning
the Kontsevich formula.

\section{Acknowledgement}

I am grateful to Patrick Delorme for many interesting discussions and
valuable comments.

\end{document}